\documentclass[nonacm,sigplan]{acmart}

\usepackage{amsmath,amsfonts}
\usepackage{physics}
\usepackage{caption}
\usepackage{qcircuit}
\usepackage{algorithm}
\usepackage{algorithmic}
\usepackage{textcomp}
\usepackage{xcolor}
\usepackage{subcaption}
\usepackage{verbatim}

\newcommand{\sol}{QuIRC}
\begin{document}

\title{Co-Designed Superconducting Architecture for Lattice Surgery of Surface Codes with Quantum Interface Routing Card} 

\author{Charles Guinn}
\authornote{cguinn@princeton.edu; samuel.stein@pnnl.gov. \textbf{These authors contributed equally to this research.}}
\affiliation{%
  \institution{Princeton University}
  \city{Princeton}
  \state{New Jersey}
  \country{USA}
}

\author{Samuel Stein}
\authornotemark[1]
\affiliation{%
  \institution{Pacific Northwest National Laboratory}
  \city{Richland}
  \state{Washington}
  \country{USA}
}

\author{Esin Tureci}
\affiliation{%
  \institution{Princeton University}
  \city{Princeton}
  \state{New Jersey}
  \country{USA}
}

\author{Guus Avis}
\affiliation{%
  \institution{University of Massachusetts Amherst}
  \city{Amherst}
  \state{Massachusetts}
  \country{USA}
}

\author{Chenxu Liu}
\affiliation{%
  \institution{Pacific Northwest National Laboratory}
  \city{Richland}
  \state{Washington}
  \country{USA}
}

\author{Stefan Krastanov}
\affiliation{%
  \institution{University of Massachusetts Amherst}
  \city{Amherst}
  \state{Massachusetts}
  \country{USA}
}

\author{Andrew A. Houck}
\affiliation{%
  \institution{Princeton University}
  \city{Princeton}
  \state{New Jersey}
  \country{USA}
}

\author{Ang Li}
\affiliation{%
  \institution{Pacific Northwest National Laboratory}
  \city{Richland}
  \state{Washington}
  \country{USA}
}

\begin{abstract}
%
Facilitating the ability to achieve logical qubit error rates below physical qubit error rates, error correction is anticipated to play an important role in scaling quantum computers. 
While many algorithms require millions of physical qubits to be executed with error correction, 
current superconducting qubit systems are only on the order of hundreds of physical qubits. 
One of the most promising codes on the superconducting qubit platform is the surface code, requiring a realistically attainable error threshold and the ability to perform universal fault-tolerant quantum computing with local operations via lattice surgery and magic state injection. 
Surface code architectures easily generalize to single-chip planar layouts due to their localized operations, however space and control hardware constraints point to limits on the number of qubits that can fit on one chip. 
Additionally, the planar routing on single-chip architectures leads to serialization of otherwise commuting gates and strain on classical decoding caused by large ancilla patches.
%
%

A distributed multi-chip architecture utilizing the surface code can potentially solve these problems if one can overcome challenges in optimizing inter-chip gates, managing collisions in networking between chips, and minimizing routing hardware costs.
We propose \emph{\sol{}}, a superconducting Quantum Interface Routing Card for Lattice Surgery between surface code modules inside of a single dilution refrigerator. \sol{} improves scaling by allowing connection of many modules, increases ancilla connectivity of surface code lattices, and offers improved transpilation of Pauli-based surface code circuits.
\sol{} employs in-situ Entangled Pair (EP) generation protocols for communication. We explore potential topological layouts of \sol{} based on superconducting hardware fabrication constraints, and demonstrate reductions in ancilla patch size by up to 77.8\%, and in layer transpilation size by 51.9\% when compared to the single-chip case. 

\sol{} will be fully open-sourced and released on GitHub. 

\end{abstract}

\maketitle 
\pagestyle{plain} 

\begin{figure}
    \centering
    \includegraphics[width=0.45\textwidth]{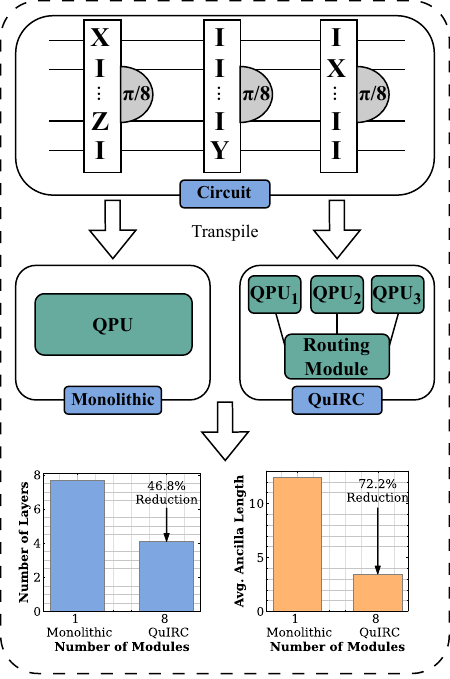}
    \caption{
    Pauli-based computations can be transpiled to single-chip or \sol{} surface code architectures. With \sol{}, the routing module increases connectivity of ancilla patches connected to logical data qubits, reducing collision in gate schedule and ancilla patch size. We find reductions in transpiled layer count by up to 46.8\% and average ancilla patch size by up to 72.2\% compared to the single-chip architecture.
    }
    \label{fig:intro_fig}
\end{figure}

\section{Introduction}\label{sec:Intro}

Quantum Computing continues to dominate headlines as a potentially revolutionizing technology \cite{moller2017impact}. 
However, for quantum computers to provide real advantage, one must surmount the overwhelming challenge of noise in quantum systems.
Currently, all the leading hardware platforms such as neutral atom \cite{wurtz2022industry}, trapped ion \cite{blinov2021comparison}, and superconducting quantum computers \cite{IBMRoadmap2023Nov}, suffer from consequential levels of noise. 
For the planar superconducting qubit platform, the focus of this work, improvements on hardware continue to be made.
Qubits are becoming increasingly coherent \cite{Place2021Mar,Wang2022Jan} and gates are becoming faster \cite{Stehlik2021Aug,Foxen2020Sep}, however the largest quantum processors today have fewer than 1000 qubits \cite{IBMRoadmap2023Nov}. 

Challenges in scaling superconducting quantum hardware include the trade-off between qubit size and coherence \cite{Wang2015Oct}, the routing of sufficient numbers of control lines to a planar chip  \cite{Krinner2019Dec}, chip packaging constraints \cite{Gao2021Nov,Huang2021Apr}, and chip yield statistics \cite{smith2022scaling}.
In addition, large-scale error correction is necessary in the development of fault-tolerant quantum computing (FTQC) as there is currently no clear pathway for physical qubits to reach noise levels low enough to implement large-scale quantum algorithms.

Quantum Error Correction (QEC) is a technique for suppressing errors in quantum systems by encoding many noisy physical qubits into a smaller number of less noisy logical qubits \cite{Preskill1998Jan}, transforming a qubit fidelity problem into a qubit scaling problem.
%
While there has been great progress in reducing resource requirements at the algorithm level \cite{Bravyi2023Aug, Heyfron2018Sep,Mosca2021Oct}, current resource estimations for implementing error corrected fault-tolerant quantum algorithms point towards requiring potentially millions of physical qubits \cite{Zhang2019Mar}.

One of the most promising error correcting codes on the superconducting platform is the surface code \cite{Kitaev2003Jan,Fowler2012Sep}.
The surface code is a Calderbank-Shor-Steane code with realistic thresholds for superconducting hardware, and requires only local operations. FTQC can be performed with local operations via lattice surgery \cite{Horsman2012Dec}.
Scaled single-chip surface code architectures have been proposed \cite{Fowler2018Aug,Litinski2019Mar}, but the physical qubit numbers required on a single chip is daunting and improving time-efficiency demands adding additional logical ancilla qubits \cite{Litinski2019Mar}. Additionally, the classical decoders' time complexity also scales with patch size \cite{Higgott2022Jul,Wu2023May}. Minimizing ancilla patch size is therefore extremely important.


One pathway to alleviate some of these constraints is through distributed quantum computing (DQC), where a quantum computer is made up of distinct nodes that share quantum communication channels.
Chips are able to host fewer qubits, trading off higher quality qubits for lower fidelity communication between nodes.
DQC architectures rely on shared entanglement that can come in the form of sharing entangled optical photons \cite{reiserer2015cavity}, microwave connections \cite{Niu2023Mar}, and mixed schemes that utilize microwave to optical conversion \cite{ang2022architectures}.
This paper specifically addresses the ideas of DQC on the planar superconducting qubit platform hosting surface code patches \cite{Litinski2019Mar}, and where multi-qubit operations are performed via lattice surgery. 
Recent technological improvements have demonstrated the feasibility of long-range microwave connections capable of performing high fidelity SWAP operations over ranges of order 1 meter inside of one dilution refrigerator \cite{Niu2023Mar, Zhong2021Feb, Leung2019Feb, Campagne-Ibarcq2018May, Axline2018Jul, Kurpiers2018Jun, Burkhart2021Aug} and longer links between refrigerators \cite{Magnard2020Dec}.
Utilizing these types of connections to share entanglement enables the departure of superconducting devices from single-chip, homogeneous "seas of qubits".
While the communication channels are expected to be noisier than qubit level noise, 
recent work demonstrated that fault tolerance can be achieved with seam error rates of $10\times$ greater than bulk error rates \cite{Ramette2023Feb}.


We propose a superconducting Quantum Interface Router Card (\sol{}) architecture that enables efficient lattice surgery operations between distributed surface code patches.
\sol{} aims to pave a road to scaled surface code architectures by reducing the need for  single-chip architectures.
It aims to also improve upon the space time trade-off for computation in square lattice surface code architectures by increasing the connectivity between surface code patches via the routing card.
We further develop a noise model to compute the effects of different routing card topologies on surface code performance. In doing so, this paper contributes:

\begin{itemize}
    \item \sol{}, a routing module for distributed lattice surgery over surface codes, improving scalability by patch size reduction and reduced transpilation overhead.
    \item A software framework for analyzing collisions in gate schedules and EP generation schedules, propagating architecture level performance metrics to algorithm level performance metrics.
    \item A reduction in average ancilla length for lattice surgery by up to 77.8\%, and a reduction in average transpiled layer counts by up to 51.9\%.

\end{itemize}

\section{Background}\label{sec:Background}
We introduce ideas in error correction and distributed quantum computing. We first discuss the surface code and how to perform Pauli-based computation with lattice surgery. We then introduce distributed quantum computing, including hardware advances and theoretical estimations for distributed surface code architectures.
\begin{figure*}
\includegraphics[width=0.95\textwidth]{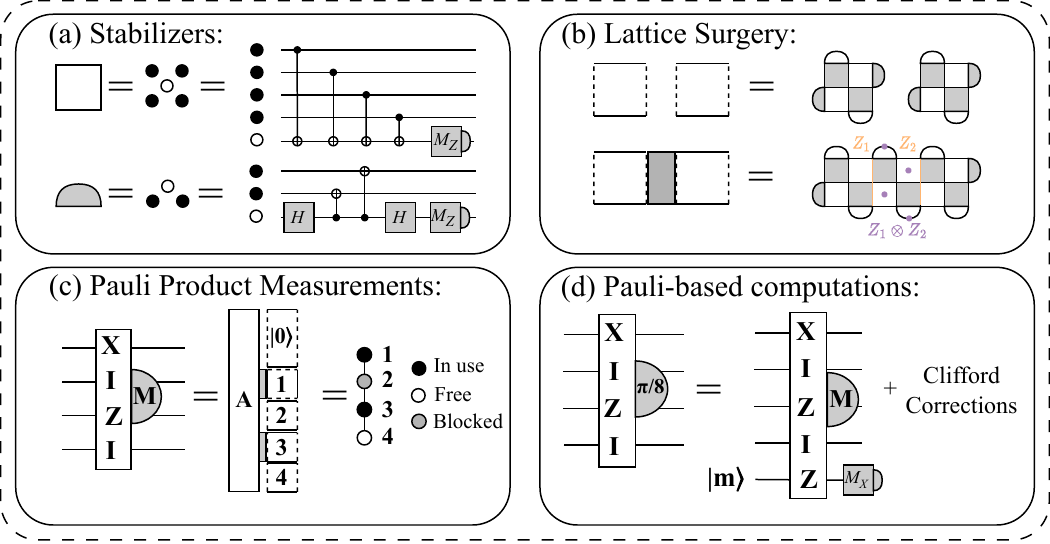}
    \caption{Surface code framework for Pauli-based circuits. (a) Stabilizer checks measure 2 or 4 qubit Pauli $X$ or $Z$ operators on data qubit plaquettes. Stabilizers are measured $d$ times to maintain fault tolerance. (b) Products of logical Pauli operators can be measured via lattice surgery by merging surface code edges to an ancilla patch. (c) Pauli product measurements of more than two operators can be measured with an ancilla patch connecting each data patch. Implementation shown on intermediate block architecture in center. (d) Universal Pauli-based  circuits can be implemented with Pauli product measurements including a magic state up to Clifford corrections.}
    \label{fig:surface code}
\end{figure*}

\subsection{Surface Code}
The patch-based rotated surface code \cite{Horsman2012Dec} encodes 1 distance $d$ logical qubit with $d^2$ data and $d^2+1$ syndrome qubits. A minimum of $d$ physical errors are required to cause a logical error. The surface code state is stabilized by alternating plaquettes of $X$ and $Z$ stabilizers that measure local parity of joint $X$ and $Z$ operators. Errors are detected by changes in the set of stabilizer measurements in each round.

The logical qubit can be initialized in $\ket{0}_L$ or $\ket{+}_L=\frac{1}{\sqrt{2}}(\ket{0}_L$ $+\ket{1}_L)$ by initializing each physical qubit in $\ket{0}$ or $\ket{+}$ and measuring each stabilizer once. 
The logical operators $X_L$ and $Z_L$ are defined on perpendicular paths on the code and are indicated by the type of stabilizer the edge is terminated with. A compact convention for drawing surface code patches is borrowed from \cite{Litinski2019Mar} and shown on the left side of Figure \ref{fig:surface code}b, where a solid edge is an $X$ or smooth edge and a dashed edge is a $Z$ or rough edge.

\subsubsection{Lattice Surgery}
Lattice surgery \cite{Horsman2012Dec} is a framework to merge and split surface code patches to measure multi-qubit operators. We introduce the simple example in Figure~\ref{fig:surface code}b to measure an operator $Z_1\otimes Z_2$. The codes are first aligned along their $Z$ edges with 1 column of data qubits between them initialized in $\ket{+}$. The merge is performed by measuring a full patch of stabilizers involving both qubits and the intermediate column. After $d$ rounds of measurement, the product of the $Z$ stabilizer outcomes between the two $Z$ operators corresponds to a fault-tolerant measurement of $Z_1\otimes Z_2$. After measurement, the codes can be split by measuring the intermediate column of data qubits and tracking corrections based on the measurement outcomes. 

This protocol can be generalized to measurements with more than two qubits as shown in Figure \ref{fig:surface code}c. In this case, the single column of extra data qubits is replaced with a full logical ancilla patch. Each qubit involved in the measurement is merged with a unique edge of the ancilla patch.

\subsubsection{Gates with Lattice Surgery}

Gates on the surface code are described with the lattice surgery framework and can be expressed naturally with Pauli based computation \cite{Litinski2019Mar, bravyi2016trading, Zhang2019Mar}. Gates are performed by a multi-patch measurement of an operator where one logical qubit is a magic state ($\ket{m}$=$\ket{0}_L+e^{i\pi/4}\ket{1}_L$), illustrated in Figure \ref{fig:surface code}d.

In hardware, Pauli-based circuits are performed by using lattice surgery merge operations to join the edges of data patches and a magic state to an ancilla patch as shown in Figure \ref{fig:surface code}d. The gate is implemented by measuring the multi-qubit Pauli operator followed by measuring the magic state. For the purposes of this work, we assume circuits expressed in an optimized Pauli-based format and focus on the problem of efficiently performing a multi-patch measurement.

\begin{figure*}
    \centering
    \includegraphics[width=\textwidth]{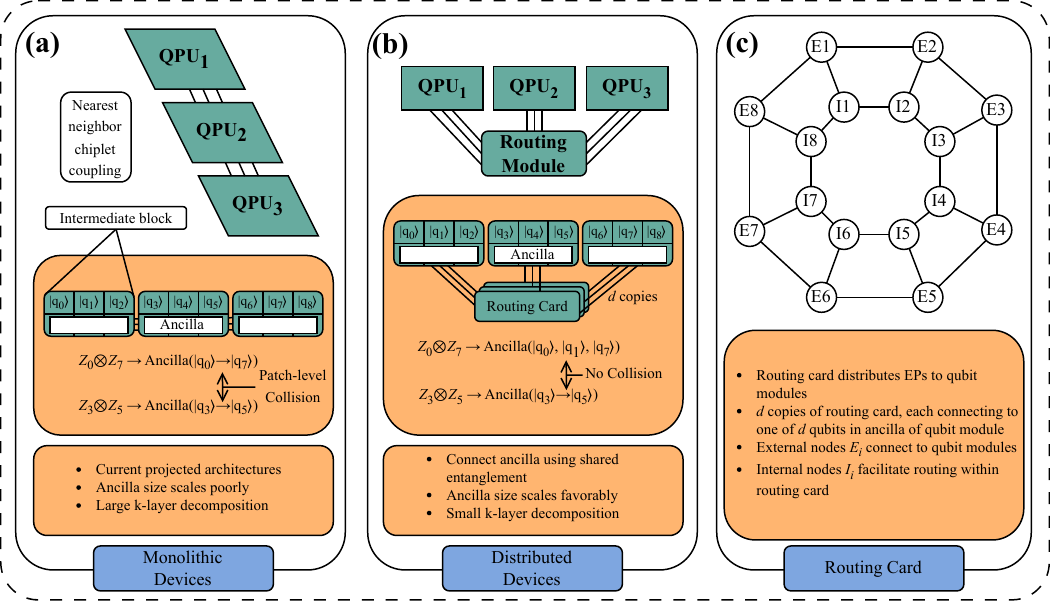}
    \caption{Distributed surface code architectures. (a) Monolithic architecture, with each QPU being an Intermediate Block. (b) Distributed architecture with routing card. (c) Potential physical layout of the routing card, with a double ring topology as an example. External nodes are denoted with $E$ and internal lodes are denoted with $I$.}
    \label{fig:int_block_layout}
\end{figure*}

\subsection{Surface Code Architectures}

\subsubsection{Single-chip Planar Architectures}
A single-chip surface code architecture is restricted to a square lattice of patches with operations on adjacent patch edges. Multi-patch measurements require layouts where ancilla patches can be routed to arbitrary edges of each data patch.  A more space-efficient layout will have a larger proportion of the ancilla patch perimeter connected to qubits participating in the measurement. Three examples of ``data blocks" are given in \cite{Litinski2019Mar} that demonstrate the space-time trade-off for patches on a square lattice. Gains in spatial efficiency can be gained by decreasing the time efficiency. Compact layouts can be achieved by allowing qubit patch rotation or adding additional ancilla as part of the measurement protocol to assist in costly $Y$ measurements. 

Another trade-off is the classical decoder dependence on the geometry of the ancilla patch used in the multi-patch measurement. An ancilla patch may need to span long distances on a chip for a measurement of far away qubits, increasing decoder complexity \cite{Higgott2022Jul}.

\subsubsection{Intermediate Block Structure}
We focus on one particular data block introduced by Litinski: the "intermediate block" \cite{Litinski2019Mar}, shown in the center of Figure \ref{fig:surface code}c.
The intermediate block is contained in a $2 \times (N+2)$ set of tiles. One row holds $N$ data qubits in a line and the other holds ancilla qubits. The extra 2 blocks on the end of the data row hold an ancilla to assist in performing $Y$ measurements. The block is called intermediate because it strikes a balance between time and space efficiency when compared to the other blocks introduced in \cite{Litinski2019Mar}. 

Each intermediate block data qubit has one edge exposed to the ancilla patch. Patch rotations can be performed to align the $X$ or $Z$ edges to the ancilla. To perform a $Y$ measurement, intermediate operations involving entanglement to and measurement of the extra 2-tile ancilla are required. Altogether, at most 5 code cycles are required to perform any multi-patch measurement. The results that follow are not unique to the intermediate block, we choose this architecture for its simple design and 1D scaling.

\subsubsection{Surface Code Connectivity as a Graph}
We reduce the intermediate block to an effective topology compatible with the routing work in upcoming sections. We first assume all intermediate steps and patch rotations have been performed ahead of time for a multi-patch measurement. This eliminates the extra ancilla from the effective topology, leaving an architecture with $N$ single tile qubits with their appropriate edges facing the ancilla patch. We also no longer have to consider particular Pauli operators in multi-patch measurements, just whether a qubit is participating in the measurement. With qubits appropriately positioned, we can consider each qubit-ancilla patch pair as a single item (a vertex), making the intermediate block a 1D line graph with $N$ vertices, shown in Figure \ref{fig:surface code}c. For magic state injections, qubits dedicated to magic state generation can be added without change to the scaling presented. We thus only consider multi-patch measurements between data qubits.

\subsection{Distributed Quantum Computing}

Single-chip superconducting quantum computing continues to face a multitude of challenges, with qubit fabrication scaling poorly with chip size \cite{smith2022scaling}, limited packaging space, and control complexity \cite{national2019quantum}.  Distributed quantum computing serves to alleviate many of these constraints \cite{ang2022architectures} at the cost of inter-chip communication complexity from transpilation \cite{wu2022autocomm,wu2022collcomm} to lower fidelity inter-chip performance \cite{ang2022architectures}. Single-fridge distributed quantum computers have already been demonstrated \cite{Niu2023Mar}, coupling four transmon chips to one central chip with $99\%$ fidelity.   

Recent theoretical demonstrations in distributed surface code show how seam error rates can tolerate $10\times p_{Bulk}$ error rates \cite{Ramette2023Feb}, a key result that motivates the use of distributed surface code architectures. With the relatively high cost of inter-chip operations, it becomes important to optimize these operations minimizing unnecessary overhead \cite{wu2022autocomm,wu2022collcomm}. 

\section{\sol{}}\label{sec:Solutions}
\sol{}s routing card architecture is motivated by advances in long-range microwave connections through superconducting coaxial cables \cite{Niu2023Mar, Zhong2021Feb, Leung2019Feb, Campagne-Ibarcq2018May, Axline2018Jul, Kurpiers2018Jun, Burkhart2021Aug}.
Cable leads can be directly wire-bonded to the edge of a chip and inductively coupled to transmon qubits on the edge of a chip \cite{Zhong2021Feb}.
Previous demonstrations have shown a 5 node quantum network with state transfer fidelity across 0.25 meter links above $F=99\%$ \cite{Niu2023Mar}.
Current trajectories envision networks of transmon chips equipped with nearest neighbor couplings connected with similar links \cite{IBMRoadmap2023Nov}.
We make use of an architecture of $N$ logical qubits divided into $M$ equal intermediate block modules from \cite{Litinski2019Mar}, based on the relatively simple block structure. When represented as a graph, each module is a line graph with a set of edges that can connect to other modules via \sol{}.

\subsection{Remote Gates on Distributed Systems}
The problem of performing a multi-patch measurement between logical qubits in intermediate block modules is primarily comprised of connecting together the appropriate ancilla patch.
In the case of a single intermediate block ($M=1$), an ancilla patch is initialized that spans between the lowest index logical qubit to the highest index logical qubit involved in the measurement. An example is shown in Figure \ref{fig:int_block_layout}a where a measurement of $Z_0 \otimes Z_7$ requires using the entire length of ancilla patch between $\ket{q_0}$ and $\ket{q_7}$. 
This can generate a large ancilla patch that contains a large number of otherwise unnecessary ancilla qubits. This results in placing substantial strain on the decoder and blocking many of the ancilla patches from being used in parallel. 

On a distributed architecture, long-range connections are used to join ancilla edges of different modules.
A $CX$ gate can be performed over the link to perform stabilizer checks the bridge the seam in the joint ancilla patch.
One way to utilize the long range connections is to use them as ``couplers'' to perform $\sqrt{SWAP}$ gates (further compiling to $CX$ gates).
This turns a distributed architecture into an effective monolithic topology (Figure \ref{fig:int_block_layout}a) while relaxing the hardware constraints on qubits per chip.

Another method is to use the cables to share EPs between modules as shown in Figure \ref{fig:graph_fig}a.
In this scheme, EPs are consumed to perform remote $CX$ gates across the seam between two modules. 
This is distinct from the coupler scheme in that the edges of modules are more flexible.
EPs can be routed to arbitrary edges while the coupler scheme has static connections. 
However, remote $CX$ gates require extra gates and measurements, which means EP-based schemes will have lower fidelity per $CX$ across the seam.

\subsection{Scaling of Microwave Links}

Scaling for the two distributed gate schemes is simple, each unit scaling linearly with $d$. We thus only consider performance of individual edge operations between physical data and ancilla qubits. For the EP based scheme, $2d$ EPs are required per round of error correction: $d$ each for the $X$ and $Z$ stabilizers.

\subsection{Intermediate Block Architectures}
We consider two architectures to perform multi-patch measurements.
The first multi-chip scheme we call monolithic and is shown in Figure \ref{fig:int_block_layout}a, where long range connections recover the $M=1$ topology of the intermediate block.
The second scheme, the routing card scheme, is shown in Figure \ref{fig:int_block_layout}b and involves a quantum routing card where modules are not directly connected to each other.
Connections between ancilla patches are instead made by passing EPs through \sol{}, the Quantum Interface Routing Card.


\subsubsection{Routing Card Architecture}
In the routing card architecture, a routing module connects each intermediate block module to other modules.
The routing module contains $d$ routing cards where each routing card connects to one qubit in each ancilla edge, illustrated in Figure \ref{fig:int_block_layout}b.
An example of a routing card with a double ring topology is shown in Figure \ref{fig:int_block_layout}c.
This architecture uses the EP based scheme for remote $CX$ gates between intermediate block modules where the path between qubits is through two cables plus qubits inside of the routing card. 

The routing card hosts physical qubits, where each edge qubit is coupled to one physical qubit in one module.
The qubits in the chip can form a graph similar to how current processors are represented, with qubits being vertices and couplings being edges.
We abide our routing architecture to realistic hardware constraints.
On the routing card, external nodes have connections to the intermediate block modules as well as local couplings, whilst internal nodes only have local couplings. This can be seen in Figure \ref{fig:int_block_layout}c.
Adding more connections between nodes decreases qubit performance, leading to a constraint on maximum degree for any node \cite{stein2023microarchitectures}.
Since edges represent physical connections via physical coupling elements, the graph should be thickness 1 on a  planar chip \cite{bravyi2023high}. However, recent proposals have introduced multi-layer chips with through-substrate vias allowing expansion to graphs of thickness 2 \cite{Grigoras2022Sep,bravyi2023high}.
\sol{} allows for arbitrary EP's to be generated between external nodes, hence noisy generation will constrain the number of nodes between any two external nodes.

\subsection{Collisions in Multi-patch Measurements}

One major issue with tiled surface code architectures is collisions in the scheduling of multi-patch measurements. When considering circuits to be made of $k$-qubit Pauli measurements \cite{bravyi2016trading,Litinski2019Mar}, ancillas must have a clear path connecting all interacting logical qubits. Even if all multi-patch measurements are made of disjoint sets of qubits and therefore commute, the ancilla patches required to connect the qubits may overlap. We refer to the type of collision caused by ancilla region overlap as a \textbf{patch-level collision}.

Executing one layered gate in the circuit, i.e. one of the Pauli measurements, requires generating an ancilla patch that touches all $k$ qubits involved in the multi-patch measurement.
In a single-chip architecture, this requires locking the entire stretch of ancilla logical qubits that span the ancilla patches in contact with the $k$ qubits involved in the measurement.
This can result in a large portion of the ancilla qubits being blocked from any other parallel use even if they are not directly involved in the Pauli measurement.

To give a probabilistic intuition into the overhead of number of ancilla patches blocked from simultaneous use we can calculate the expected span of randomly selected $k$ qubits on an $N$ qubit single-chip architecture. The expectation value of the largest qubit index $k_{max}$ is:
\begin{equation}
    \mathbb{E}[k_{max}] = \frac{k \cdot (N+1)}{(k+1)}
\end{equation}
and we can write the smallest qubit index using symmetry as:
\begin{equation}
    \mathbb{E}[k_{min}] = N - \frac{k \cdot (N+1)}{(k+1)}
\end{equation}
giving the expectation value of the span of randomly selected $k$ qubits as:
\begin{equation}\label{eqn:bins}
    \mathbb{E}[k_{span}] =  \frac{2 \cdot k \cdot (N+1)}{(k+1)} - N \\
    = \frac{N \cdot (k-1) + 2k}{k+1}
\end{equation}
which grows linearly with $N$.
In a distributed architecture with $M>1$ qubits per chip, the total ancilla patch size is reduced to grow with $N/M$ instead. 
A specific example is the comparison between the monolithic and distributed cases in Figure \ref{fig:int_block_layout}a-b.
In the monolithic architecture the ancilla patch spanning $\ket{q_0}$ to $\ket{q_7}$ includes $\ket{q_3}$ and $\ket{q_5}$ leading to a collision in measuring $Z_0 \otimes Z_7$ and $Z_3\otimes Z_5$ simultaneously. 
In the distributed device, the connection through the routing card allows the patch for $Z_0 \otimes Z_7$ to exclude $\ket{q_3} \rightarrow \ket{q_5}$ leading to no collision. A general circuit schedule can be transpiled into a schedule of circuit layers on our architecture.


\begin{figure}
    \centering
    \includegraphics[width=0.5\textwidth]{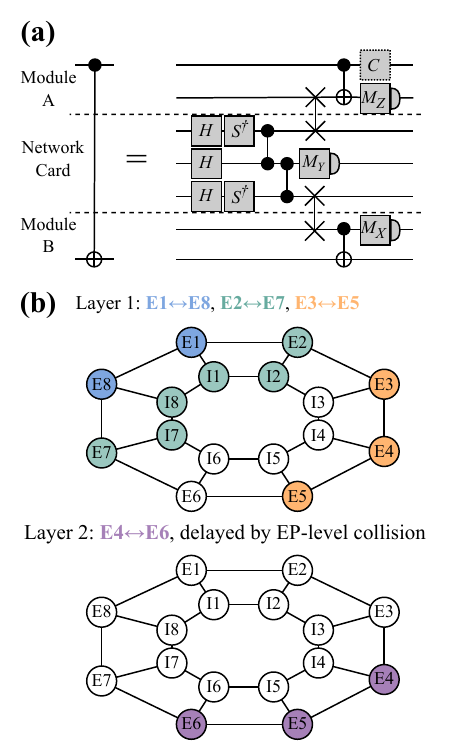}
    \caption{Routing card operation. (a) Remote EP generation plus remote $CX$ execution for stabilizer circuit. The measurements determine the Pauli correction $C$ to the final operation. It does not need to be applied, just absorbed into future measurement outcomes. (b) Example of an EP generation schedule broken down into two layers due to EP-level collisions in the routing card.}
    \label{fig:graph_fig}
\end{figure}

\subsection{Scheduling Entangled Pair Generation}\label{sec:EP gen}
Given a valid layer decomposition for a circuit schedule, there is additional possibility of latency due to EP generation traffic. One circuit layer requires all EPs to arrive at each module for each round of error correction. A realistic routing card will not have all-to-all connectivity between external nodes and may thus have collisions in a sufficiently dense EP schedule. Before studying this, we introduce our EP generation scheme via linear graph states.

\subsubsection{Generating EPs}

The purpose of the routing card is to perform entanglement switching~\cite{vardoyan2021a, avis2023, coopmans2021, dai2021, panigrahy2022}, that is, to create entanglement between arbitrary pairs of qubits in different modules.
Each round of error correction calls a set of EPs that needs to be generated within the routing card and distributed to the external nodes connected to the proper modules.
Preparing an EP starts with preparation of a linear graph state \cite{hein2006} between the external nodes of interest along a path through the qubits in the routing card.
Performing $Y$ basis measurements of the qubits in the graph state between the external nodes performs local complementations up to Pauli corrections resulting in a two-qubit entangled state between the external modes in constant time. 

To prepare the Bell state $\ket{\Phi^+}$ on the external nodes, the graph state preparation is modified. 
We perform the following protocol for a path of length $\nu$ connecting two external nodes:

\begin{itemize}
    \item Prepare qubits 1 through $\nu$ in the state $\ket +$.
    \item Apply $S^\dagger$ to qubits 3 through $\nu$, where $S$ is the $S$ gate.
    \item At qubit $1$, apply $(S^\dagger)^{\nu-2}$; this will be $I$, $S^\dagger$, $Z$ or $S$ depending on the value of $\nu - 2 \mod 4$.
    \item Perform $CZ$ gates between qubits $i$ and $i + 1$ for $i = 1, 2, ... \nu-1$, followed by a Hadamard gate on qubit $\nu - 1$.
    \item Measure the qubits 2 through $\nu-1$ in the $Y$ basis and denote the outcome of qubit $i$ as $m_i = 0, 1$.
    \item At qubit $1$, track the correction $C$
    \begin{equation}
    C = Z^{\sum_{i=2}^{\nu-1} (\nu - i) m_i} X^{\sum_{i=2}^{\nu-1} m_i}.
    \end{equation}
    \item SWAP the states from qubits 1 and $\nu$ to the modules. Because the correction $C$ is a Pauli correction, it can be absorbed into the final measurement outcomes of the remote gate and stabilizer measurement.
\end{itemize}

With an EP shared between the modules, a remote $CX$ gate can be performed with only local operations between the two modules. The full circuit for a round of $EP$ generation and remote gate execution is shown in Figure \ref{fig:graph_fig}a. 

We note that the above protocol only requires the use of each edge and measurement of each qubit along the path once, and the measurements can be performed in parallel.
However, when multiple pairs of remote qubits need to be connected at the same time, collisions will occur if there does not exist an appropriate set of node-disjoint paths in the routing-card graph.
We call this type of collision in the routing card \textbf{EP-level collision} with an example shown in Figure \ref{fig:graph_fig}b.

There are cases where such collisions could be avoided by utilizing a smarter protocol, for instance with a cleverly chosen graph state that includes multiple pairs of external nodes \cite{hahn2018}.
However, there are limits on how well collisions can be avoided using even an optimal protocol based on the ability to disconnect the graphs of desired EPs via a certain number of edges.
This follows from the fact that a quantum network can be used to simulate the routing card through entanglement-mediated gates~\cite{eisert2000}, and limits on what can be achieved on quantum networks are known~\cite{pirandola2019}.
For example, in a dumbbell graph where two groups of two external nodes are connected through one edge, it is impossible to create two entangled pairs between those sets using the bottleneck edge only once, thereby introducing extra latency or collisions.
This shows that even though the protocol we propose may not always be the optimal protocol, there are clear limits to how much it could be improved.

\section{\sol{} Experimental Settings and  Results}\label{sec:Benchmarks}
In this section we present the experimental settings for \sol{}, discuss how simulation parameters are attained, and demonstrate \sol{}s performance. 

\subsection{Experiment Settings}

To evaluate \sol{}s architecture, we make use of Python and packages NetworkX, Qiskit \cite{Qiskit}, and STIM \cite{Gidney2021Jul}. We set the total number of logical qubits in all simulations to be $N=24$, distributed across $M$ modules. For simulating ancilla length and layer decomposition, we sample 100 random sets of $P$ $K$-qubit operators and report the means. We restrict $P*K=24$ to study fully dense schedules. We make use of a brute force search to search the traveling salesman solution for shortest path both for ancilla patch and EP routing.

\begin{figure}
    \centering
    \includegraphics[width=0.5\textwidth]{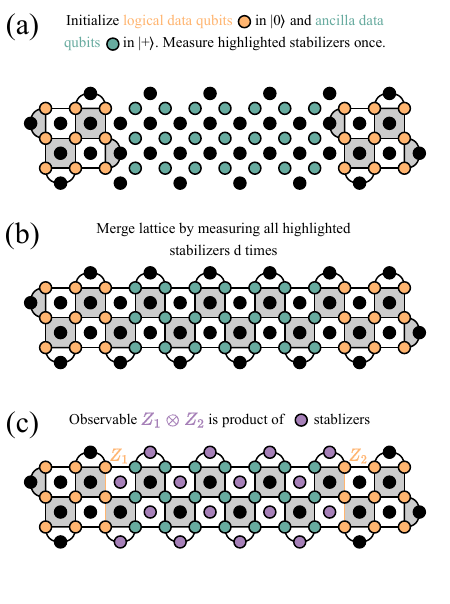}
    \caption{Physical layout for STIM simulations, $d=3$. (a) Initialization step prepares logical qubits in $Z$ basis. White and gray plaquette stabilizers are measured once. (b) Lattice surgery merge creates joint ancilla patch. White and gray plaquette stabilizers are measured $d$ times. (c) Logical operator $Z_1 \otimes Z_2$ is given by product of purple stabilizer outcomes.}
    \label{fig:LS_cartoon}
\end{figure}

To simulate logical error rates in STIM, we prepare distance $d$ logical qubits separated by a size $2d+1$ ancilla patch. After initializing logical qubits in the $Z$ basis, lattice surgery is implemented in the $X$ basis by measuring the full set of stabilizers across the entire joint data-ancilla-data patch, illustrated in Figure \ref{fig:LS_cartoon}. Noise is injected according to the model in Section \ref{sec:error_rates}. The PyMatching package \cite{Higgott2022Jul} is used to decode $d$ rounds of stabilizer measurements and detect logical errors in the observable $Z_1 \otimes Z_2$ averaged over 10 million shots.

\subsection{Collisions in Distributed Intermediate Block}
To evaluate the performance of the distributed intermediate block, we consider the trade-off of ancilla patch economy versus latency for EP distribution.

\begin{figure}
    \centering
    \includegraphics[width=0.5\textwidth]{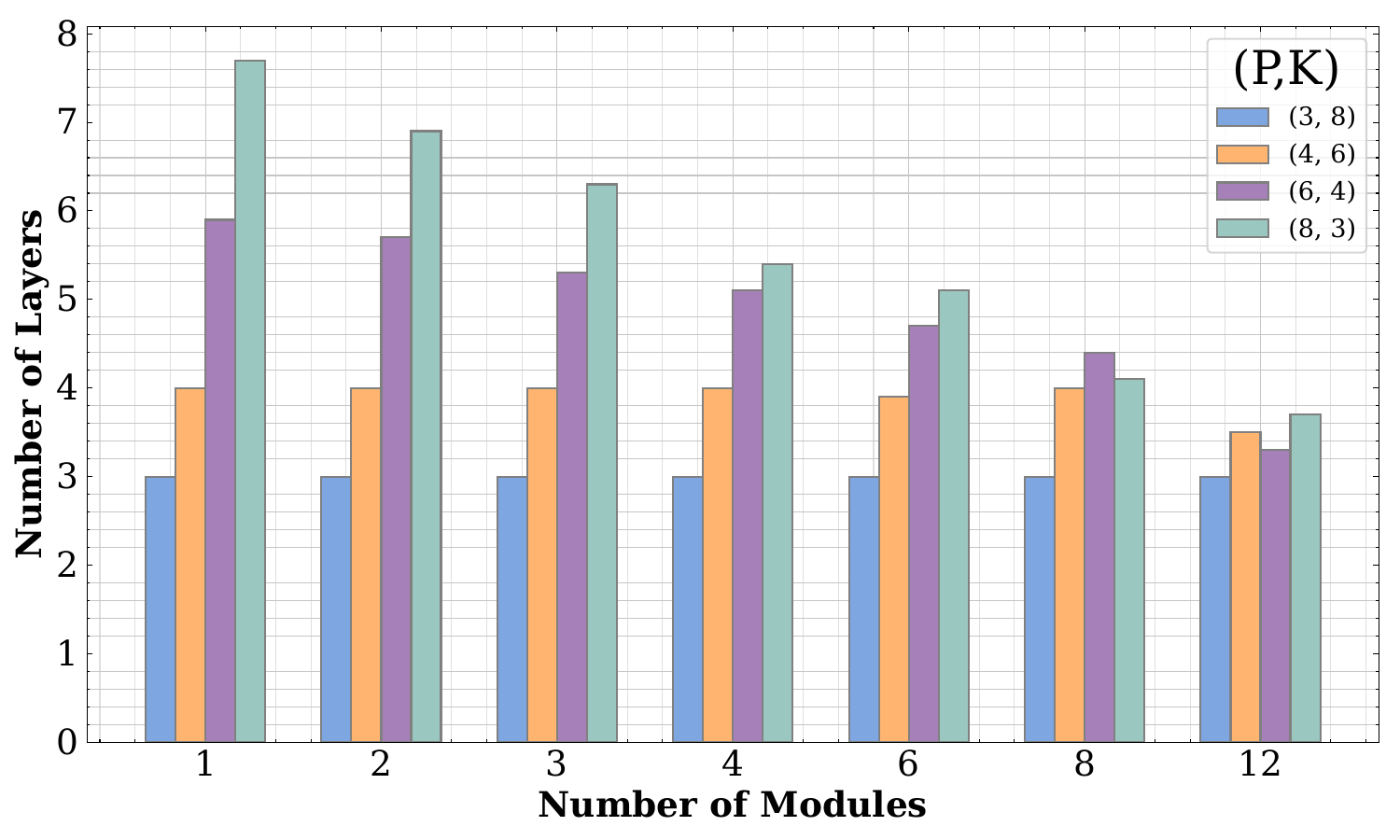}
    \caption{ Scheduling benchmarks for layer transpilation over different (P,K) operators against varying number of modules.}
    \label{fig:Patch Layers}
\end{figure}
\begin{figure}
    \centering
    \includegraphics[width=0.5\textwidth]{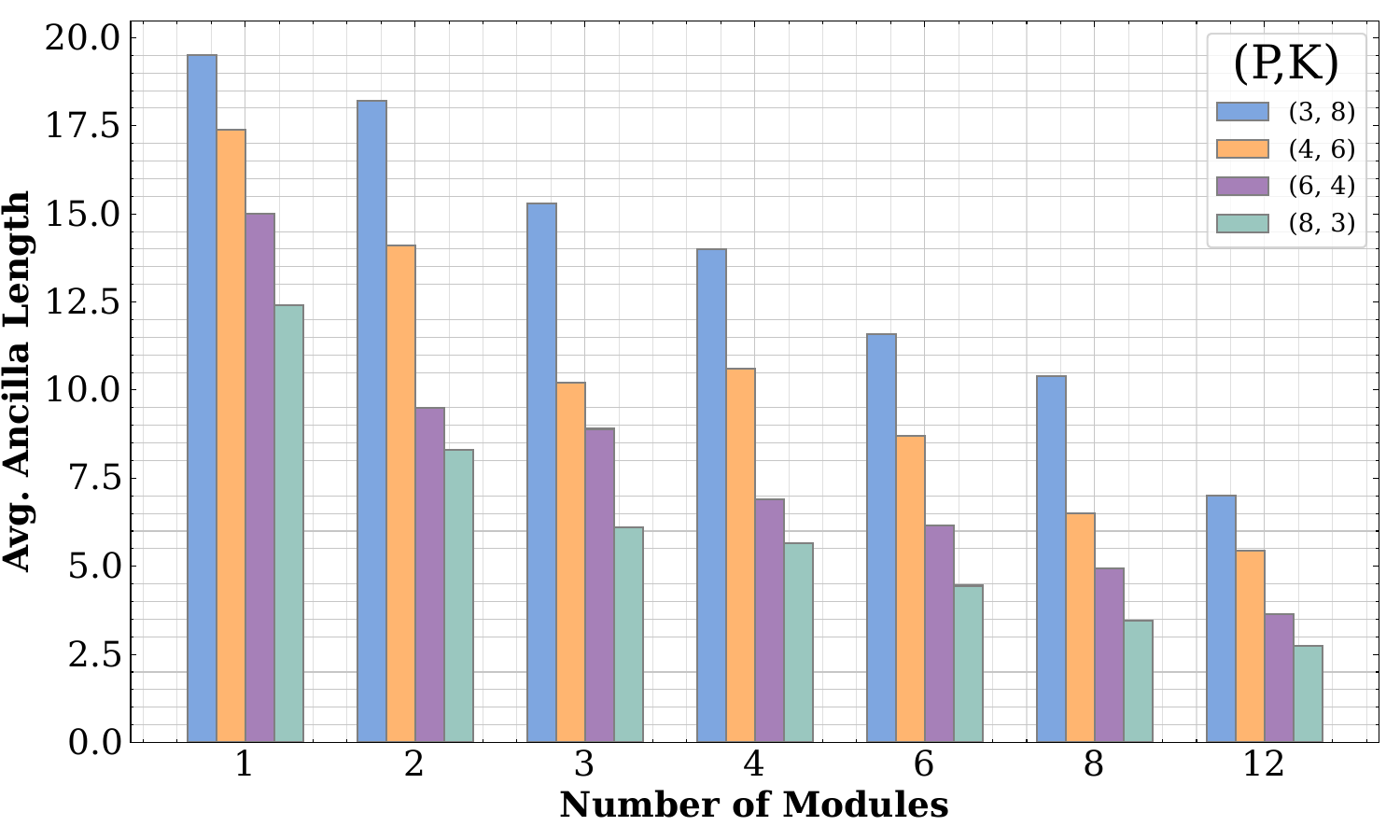}
    \caption{ Scheduling benchmarks for ancilla length over different (P,K) operators against varying number of modules.}
    \label{fig:Patch Size}
\end{figure}
\begin{figure}
    \centering
    \includegraphics[width=0.5\textwidth]{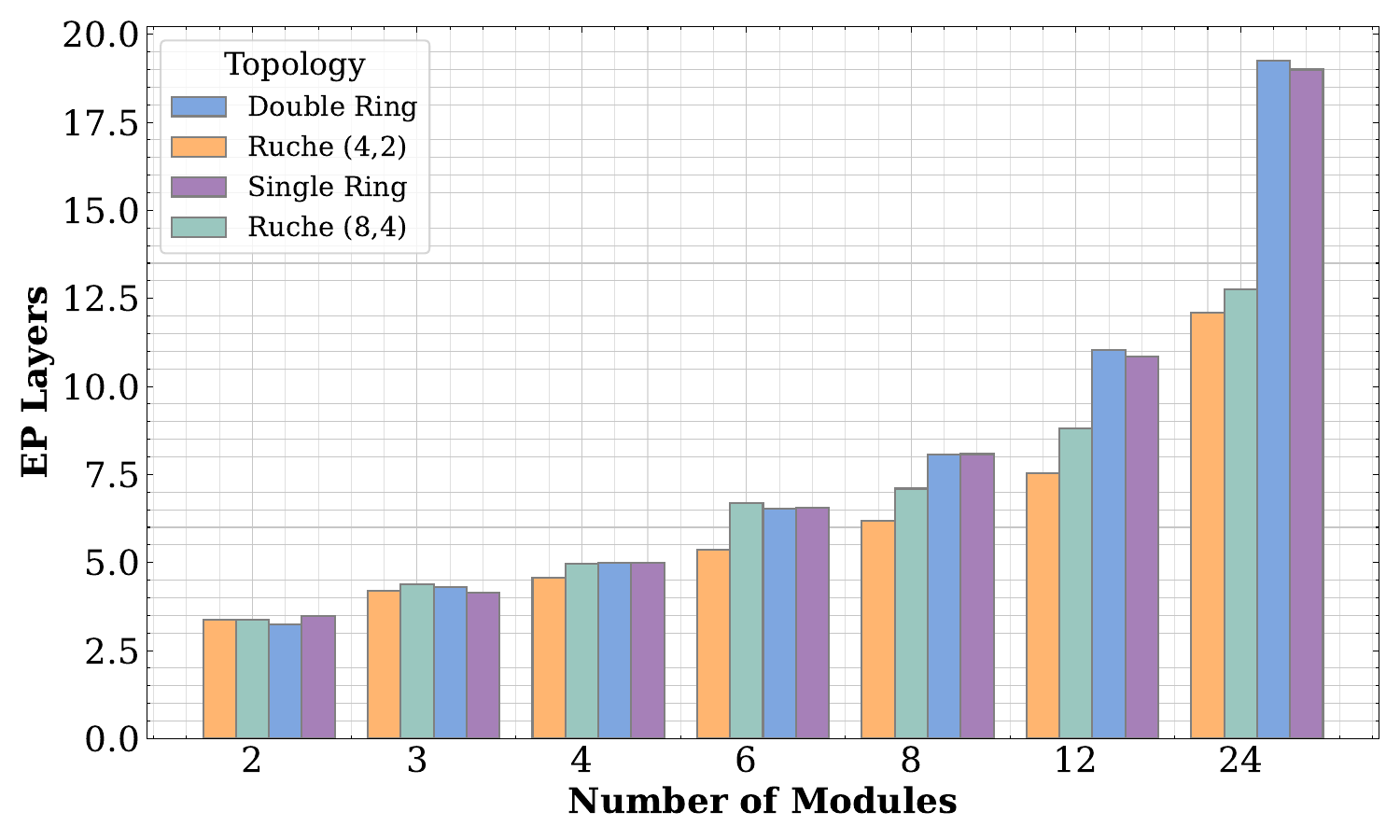}
    \caption{ Scheduling benchmarks for EP Layers over different routing card topologies over varying number of modules.}
    \label{fig:EP_Layers}
\end{figure}


\subsubsection{Patch-level Collisions}
To start with ancilla patch economy, we consider random maximum density circuit slices.
For $P=N/K$ simultaneous $K$-qubit multi-patch measurements, we determine how many layers of multi-patch measurements are required to execute every gate.
To capture only latency given by patch-level collisions, we assume the length $L=N/M$ intermediate block modules are connected to routing card containing external nodes connecting at indices $L/3$ and $2L/3$. 
To simulate this, we generate random sets of $P$ $K$-qubit operators.
We then perform a traveling salesman search over each operator in $P$, constrained by only searching paths that move forward. 
If a path exists, the operator is removed from the sample and the path is removed from the graph, otherwise the operator remains in the sample. Once all operators have been checked for a valid path, the graph is reinitialized, representing a new layer, and the search begins again over the remaining sample. This is repeated until no operators are left, and gives the total number of layers required to execute a sample of $P$ $K$-qubit operators. For ancilla length, we simply keep track of the average length of the paths discovered from the traveling salesman search.

In the extreme distributed case ($M=N$)  all $P$ measurements can be executed in exactly 1 layer because no extra ancilla patches are required to bridge between qubits.
On the other hand, the worst case single-chip circuit slice needs $P$ layers to execute.
An example of this is a schedule of measuring two-body operators where qubit indices are paired as ${i,N+1-i}$. 
Measuring $Z_1 \otimes Z_N$ requires an ancilla patch that spans the whole module, and so on for increasing $i$.
On average, there will be some operators that can always be measured simultaneously, and increasing $M$ allows more pathways between qubits further decreasing layer count.
This is illustrated in Figure \ref{fig:Patch Layers}.

The best results are seen for schedules involving a large number of few-qubit operators.
For $P=8,K=3$ there is a 1.92x decrease in the number of layers for the circuit slice on average.
The reason for this is ancilla patches for few-qubit measurements are more likely to occupy disjoint module sets if there are more modules.
For the case of $P=3,k=8$ on the other end of the spectrum, there is no improvement when increasing number of modules because it is unlikely that any one measurement lives in a module disjoint from the other two.

As the number of modules increase, the decrease in number of layers is a direct consequence of decrease in the number of ancilla patches that needs to be used in each layer. Figure \ref{fig:Patch Size} shows average ancilla size for $K$-qubit operators as a function of module size.
\emph{For all $K$, increasing the number of modules decreases the total ancilla patch length.} The ancilla size reductions from 1 to 12 modules range from $6.67\%$ to $77.82\%$ depending on $P$ and $K$.



\begin{table}[!t]
\centering
\caption{Average relative percentage reductions for layer decomposition and ancilla size over different (\$P\$,\$K\$) combinations as compared to the single-chip case.}
\begin{tabular}{|c|c|c|c|c|}
\hline
& (3, 8) & (4, 6) & (6, 4) & (8, 3) \\ 
\hline
\multicolumn{5}{|c|}{Averagae Layer Decomposition Improvements} \\
\hline
2 & 0.0\% & 0.0\% & 3.39\% & 10.3\% \\
3 & 0.0\% & 0.0\% & 10.17\% & 18.1\% \\
4 & 0.0\% & 0.0\% & 13.56\% & 29.8\% \\
6 & 0.0\% & 2.5\% & 20.34\% & 33.7\% \\
8 & 0.0\% & 0.0\% & 25.42\% & 46.7\% \\
12 & 0.0\% & 12.5\% & 44.07\% & 51.9\% \\
\hline
\multicolumn{5}{|c|}{Average Ancilla Size Improvements} \\
\hline
2 & 64.1\% & 68.6\% & 75.6\% & 77.8\% \\
3 & 46.6\% & 62.6\% & 67.0\% & 72.1\% \\
4 & 40.5\% & 50.0\% & 59.0\% & 64.1\% \\
6 & 28.2\% & 39.0\% & 54.0\% & 54.4\% \\
8 & 21.5\% & 41.3\% & 40.6\% & 50.8\% \\
12 & 6.6\% & 18.9\% & 36.6\% & 33.0\% \\
\hline
\end{tabular}
\label{tab:combined_table}
\end{table}

\subsubsection{EP-based Collisions}
For a given layer schedule that is valid for a routing card architecture, the question remains of how best to distribute the EPs.
Each module involved in a layer will use up to 2 external nodes in the routing card that need to receive a half-EP to execute a layer.
If more EPs are required for a layer, the odds of collisions in the routing card are increased.
This introduces a balance in choosing $M$, where increasing $M$ decreases the total number of layers in a circuit slice at the cost of layers taking longer to execute.

To benchmark this, we consider random full density EP requests on a routing card.
An $M$ module routing card will have $2M$ external nodes to enable proper ancilla chain generation.
One layer will consist of $M$ total EPs distributed between the external nodes.
This choice to use uniform, full density circuit layers provides intuition into what cases a distributed architecture is beneficial, and we note real algorithms will involve less demand on the routing card.

We consider four topologies of our routing card, categorized by thickness. For graph thickness 1, we consider ring and double-ring topologies.
We also consider the graph thickness 2 Ruche networks \cite{jung2020ruche} Ruche(4,2) and Ruche(8,4). Ruche($i,j$) networks topologies are ring networks where every $j$-th qubit has an additional connection to the qubit $i$ indices ahead of it.

Figure \ref{fig:EP_Layers} plots the average number of EP generation layers required to generate the full set of $M$ EPs.
We observe two key takeaways.
First, \emph{more modules leads to more latency because the EP schedule is larger.}
Second, \emph{the thickness 2 graph topologies significantly outperform the thickness 1 graph topologies for high module numbers.} 
This is expected with higher connectivity networks, however we further emphasize the need to optimize through-substrate via technology \cite{Grigoras2022Sep} to enable thickness-2 graphs. 

\subsection{Surface Code Simulations}

In this section, we detail how we model distributed lattice surgery results and present our results. 

\subsubsection{Surface Code Modeling}
Performing a multi-patch measurement on distributed surface code patches requires an in-homogeneous noise model.
Local and remote gates will have different fidelities and scheduling constraints with the routing card will introduce latency which a noise model must capture.
We benchmark the threshold for successful measurement of a multi-qubit operator in the presence of a noisy seam and latency introduced by the routing card.

\begin{algorithm}[!t]
\caption{\sol{} Simulation Model. $Q_L$ represents a logical qubit, $Q_{L,Data}$ represent physical data qubits a logical patch, and $Q_{L,Syndrome}$ are physical syndrome qubits in a logical patch. $p_{\text{Error Type}}$ applies a depolarizing channel of that probability. $Stabilizer(Q_L)$ performs gates for syndrome extraction on patch $Q_L$. $Measure(Q_{L})$ measures syndrome qubits in $Q_L$. $Merge({Q})$ defines a new logical patch from the set ${Q}$.}
\label{alg:master}
\begin{algorithmic}
\STATE \text{Initialize physical qubits}
\STATE $Q_{L1},Q_{L2} \gets \text{Data Logical Qubits}$
\STATE $Q_{A} \gets \text{Logical ancilla patch}$
\STATE $Q_{L1} \gets \ket{0} \text{ or } X\ket{0} \text{ with } p_{SPAM} $
\STATE $Q_{L2} \gets \ket{0} \text{ or } X\ket{0} \text{ with } p_{SPAM} $
\STATE $Q_{A} \gets \ket{+} \text{ or } Z\ket{+} \text{ with } p_{SPAM}$
\STATE \text{Initialize logical data qubits}
\STATE $Q_{L1} \gets Stabilizer(Q_{L1}) \text{ with } p_{local}$
\STATE $Q_{L2} \gets Stabilizer(Q_{L2}) \text{ with } p_{local}$
\STATE $M_{1}\gets Measure(Q_{L1}) \text{ with } p_{SPAM}$
\STATE $M_{2}\gets Measure(Q_{L2}) \text{ with } p_{SPAM}$
\STATE \text{Merge Logical Data Qubits Over Ancilla Patch}
\STATE $Q_{LS} \gets Merge(Q_{L1},Q_{A},Q_{L2})$
\STATE $Q_{LS} \gets p_{Latency}$
\FOR {\text{round in d rounds }}
    \STATE $Q_{LS} \gets Stabilizer(Q_{LS}) \text{ with } p_{local}$
    \STATE $Q_{LS} \gets p_{Remote} \text{ if remote seam} $
    \STATE $M_{LS} \gets Measure(Q_{LS}) \text{ with } p_{SPAM}$
\ENDFOR
\STATE $\text{Observe Stabilizers} = \prod M_{LS}$

\end{algorithmic}
\end{algorithm}

\subsubsection{Error Rates}\label{sec:error_rates}
We define the following error rates for simulations:

\begin{itemize}
    \item $p_{SPAM}$: Error from state preparation and measurement. Assumed to be $X$ or $Z$ error depending on the basis of preparation or measurement.
    \item $p_{local}$: Local gate error for Hadamard ($H$) and in-module $CX$ gates. Assumed to be depolarizing noise per gate.
    \item $p_{remote,X}(N),p_{remote,Z}(N)$: Remote $CX$ gate errors. We assume this to depend only on the number and fidelity of measurements required to perform the remote gate, valid for $p_{SPAM}>>p_{local}$. Measurements during the EP generation can apply $X$ and $Y$ errors depending on $N$ while the $Z$ and $X$ basis measurements can apply $X$ or $Z$ errors on the target and control qubits respectively.
    \item $p_{latency}$: Error accrued while waiting for routing card to distribute all EPs for a $CX$ layer in an error correction cycle. Assumed to be depolarizing noise with magnitude a function of number of circuit steps waited and error per circuit step.
\end{itemize}

In superconducting systems, $p_{SPAM}$ is generally limited to order $1\%$ by measurement time \cite{Heinsoo2018Sep}.
In this work we will consider the preparation and measurement errors to be fixed at $p_{SPAM}=0.01$. 
The local gate error $p_{local}$ is often dominated by two-qubit gate errors.
Scaled systems have average error rates around $1\%$ with smaller demonstrations approaching $0.1\%$ \cite{Foxen2020Sep,Stehlik2021Aug}.
The local gate error is the main parameter we will sweep to find surface code thresholds.


\begin{figure*}
    \centering
    \includegraphics[width=0.971\textwidth]{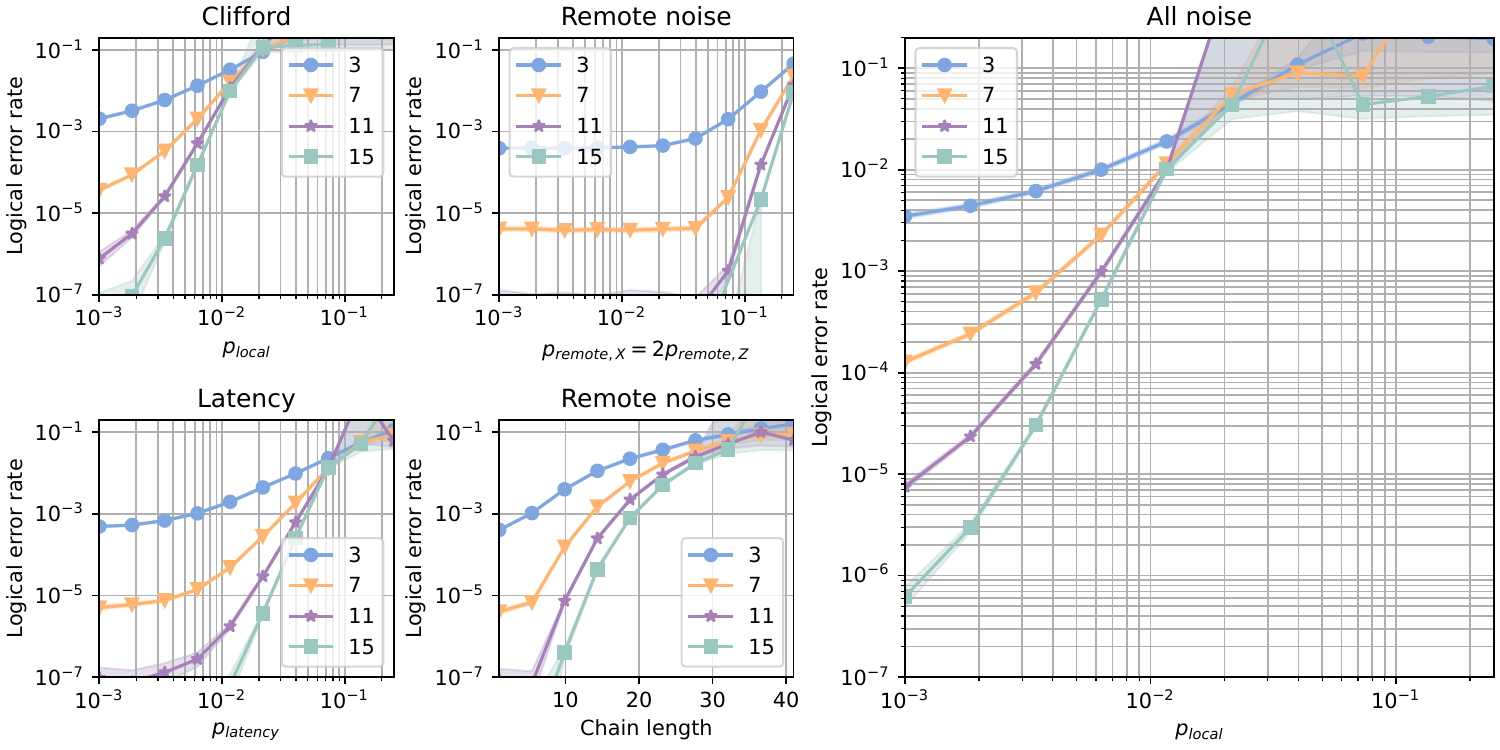}
    \caption{(color online) Surface code performance for different error models. Logical error rates are for two-qubit $Z \otimes Z$ measurements with lattice surgery. $p_{SPAM}$ is fixed at 0.01 resulting in plateaus in trajectories for low error. (Clifford) Measurement with only local gate and measurement noise. (Remote Noise - Upper) Measurement with additional 5\% error on remote gate operations. (Latency) Measurement with additional 1\% depolarizing noise from routing card latency. (Remote Noise - Lower) Measurement with additional remote gate and latency errors.}
    \label{fig:surfacecode_results}
\end{figure*}

The remote gate errors $p_{remote,X},p_{remote,Z}$ are the probabilities of a $X$ and $Z$ errors on the target and control qubits for the remote $CX$.
During the EP generation, $X$ errors propagate to the target qubit and $Y$ errors separate into $X$ and $Z$ errors on different qubits.
Additionally, the extra $X$ and $Z$ basis measurements in the remote gate execution give additional $Z$ and $X$ errors.
We approximate the EP generation error probabilities for $X$ and $Y$ as $\nu*p_{SPAM}/2$, leading to the expressions:

\begin{equation}\label{eqn:chain_length}
\begin{split}
    p_{remote,X} \approx \nu(p_{SPAM}),  \\
    p_{remote,Z} \approx \frac{\nu}{2}(p_{SPAM})
\end{split}
\end{equation}

Lastly, we estimate $p_{latency}$ by examining the $EP$ generation circuits.
One $EP$ generation via a graph state takes one single-qubit gate layer, two $CZ$ layers, one $SWAP$ layer to move the EPs from the routing card to the modules, one local $CX$ layer in the modules, and a measurement layer.
With typical gate and measurement times of $T_{gate}=100 ns$ and $T_{meas}=1$ $\mu$s respectively, we give a conservative estimate of $T_{EP}=2$ $\mu$s for a single round of EP generation.
Qubit coherence times vary but can be reliably lower bounded by $T_1=100\mu s$ \cite{Place2021Mar,Wang2022Jan}.
Overall, this gives an effective upper bound on $p_{latency}=\exp(-T_{EP}/T_{1})=0.02$. Using more optimistic (but still reasonable) parameters of $T_{EP}=1$ $\mu$s and $T_1=1$ ms, a latency error of $p_{latency}=0.001$ is achievable. 

 A single measurement with error rates applied is described in Algorithm \ref{alg:master}. 
 

\subsection{Surface Code Simulation}
We now quantify how additional errors from a routing card affect surface code properties. The small plots in Figure \ref{fig:surfacecode_results} show the logical error rates for each parameter in our noise model in isolation. 
We fix $p_{SPAM}=0.01$ and report thresholds for $p_{local}$, $p_{remote,x}=2p_{remote,z}$, and $p_{latency}$ of approximately 2\%, 20\%, and 10\% respectively. 
The thresholds for $p_{local}$ and $p_{remote}$ are in agreement with literature \cite{Fowler2012Sep,Ramette2023Feb, Fowler2010May}. 
Because we do not include $p_{SPAM}$ in the error being varied, we observe plateaus in logical error rates when $p_{SPAM}$ becomes the dominant error. 

We may also interpret these results in the context of a distributed system relying on EP generation. Figure \ref{fig:surfacecode_results} expresses remote noise via Equation \ref{eqn:chain_length} to extract error rates based on chain length $\nu$ in the routing card. Additionally, the number of EP layers for a gate layer can be converted to a latency error through qubit coherence time and EP generation time. 

The results in Figure \ref{fig:EP_Layers} demonstrate that it is possible to operate \sol{} below the latency threshold. 
For a modest number of modules, the number of EP layers in the strictest scheduling case remains below 10 leading to $p_{latency}\approx 0.01$ for realistic hardware. 
For a sufficiently connected routing card topology such as the Ruche network, the length of connecting paths in the routing card also remains low making it possible to operate \sol{} below the remote noise threshold. 
We simulate the logical error rate for a full noise model in Figure \ref{fig:surfacecode_results} with the choice $\nu=6$ and $p_{latency}=0.01$. Compared to the local-only noise model representing the single-chip case, the full noise model has similar threshold and approximately one order of magnitude higher error rate. 
This shows that a \emph{\sol{} architecture can achieve similar code performance to the single-chip case even in the presence of errors due to networking.} 
This is in addition to the circuit-level layer transpilation improvements from Figure \ref{fig:Patch Layers} and ease in fabrication and packaging afforded by smaller modules. 







\section{Related Work}

In the field of surface code architecture, a large amount of work involves efficiently compiling quantum circuits and magic state resources onto single-chip surface code architectures \cite{Litinski2019Mar, Lao2018Sep,Paler2017Apr,Herr2017Jan, Holmes2019Jun}. In other words, the physical architecture is fixed and the problem is partitioning square tiles efficiently. Our work optimizes along a different axis, aiming to improve the hardware connectivity itself.

Of special note is Litinski's distributed architecture \cite{Litinski2019Mar} where gates can be done time-optimally \cite{Fowler2012Oct} using quantum computers that can share EPs. This was shown to help parallelize circuits where each module contained qubits equal to the size of the circuit whereas our work separates the full set of qubits between modules.

With respect to networking and interface cards, much of the literature focuses on photonic quantum computing, and makes use of quantum repeaters \cite{li2019experimental,li2023all,azuma2015all}, which are used to perform long range entanglement generation. These papers often address communication between nodes, but do not look at how to schedule and mediate lattice surgery over a network. 

Other works in superconducting network scheduling look to minimize inter-node communications \cite{wu2022autocomm, wu2022collcomm, Wu2023May}. This research however looks at generating a transpilation optimization technique that minimizes the use of the inter-node gates. This is orthogonal to our work, as we employ a repeated fixed structure of communication between nodes to perform lattice surgery. 
\section{Conclusion}

In this paper, we presented \sol{}, a superconducting quantum interface routing architecture for distributed surface code systems.
When comparing our architecture to the single-chip equivalent, we demonstrate improvements up to 77.8\% in average ancilla size, and up to 51.9\% reductions in  layer transpilation depth. This substantially reduces decoder strain and improves transpiled circuit depths. The \sol{} architecture is guided by current hardware constraints, and design is soft-constrained around these factors.

In doing this analysis, we developed a surface code noise model compatible with Stim that captures the essence of errors due to measurement based EP generation, remote gate execution, and EP-level latency, allowing for rapid exploration of various \sol{} architectures. Different hardware noise structures can be directly translated to logical error rates and constraints on routing card topologies.

Trade-offs between patch-level and EP-level latency for a $24$ logical data qubit system as a function of module number have been demonstrated, highlighting the need to strike a balance between high numbers of modules with fewer patch-level collisions and small numbers of modules with reduced EP traffic. Our results are representative of the most demanding benchmarks of Pauli product measurement protocols with lattice surgery, with layers being fully occupied by Pauli operators.

This work provides a building block in the distributed superconducting error correction architecture domain, however leaves much room for optimization and alternative investigation. Future work may expand our results to other data block architectures and include magic state distillation for exploring fully universal architectures. Another future direction would be to compile specific Pauli-based computation circuits to a distributed architecture, opening the door to co-designed architectures for specific applications. Furthermore, other problems in topology, transpilation, and entanglement protocol selection all remain open, and should be investigated.

\section*{Acknowledgements}
We would like to thank Joshua Ramette, Henry Prestegaard, and Shashwat Kumar for their insightful discussions. This material is based upon work supported by the U.S. Department of Energy, Office of Science, National Quantum Information Science Research Centers, Co-design Center for Quantum Advantage (C2QA) under contract number DE-SC0012704. This work is partly funded by EPiQC, an NSF Expedition in Computing, under grants CCF-1730082/1730449. The Pacific Northwest National Laboratory is operated by Battelle for the U.S. Department of Energy under Contract DE-AC05-76RL01830. SK and GA are supported by the CQN under NSF grant 1941583.

\newpage
\bibliographystyle{plain}
\bibliography{refs}

\begin{thebibliography}{10}

\bibitem{ang2022architectures}
James Ang, Gabriella Carini, Yanzhu Chen, Isaac Chuang, Michael~Austin DeMarco, Sophia~E Economou, Alec Eickbusch, Andrei Faraon, Kai-Mei Fu, Steven~M Girvin, et~al.
\newblock Architectures for multinode superconducting quantum computers.
\newblock {\em arXiv preprint arXiv:2212.06167}, 2022.

\bibitem{avis2023}
Guus Avis, Filip Rozp{\k{e}}dek, and Stephanie Wehner.
\newblock Analysis of multipartite entanglement distribution using a central quantum-network node.
\newblock {\em Physical Review A}, 107(1):012609, January 2023.

\bibitem{Axline2018Jul}
Christopher~J. Axline, Luke~D. Burkhart, Wolfgang Pfaff, Mengzhen Zhang, Kevin Chou, Philippe Campagne-Ibarcq, Philip Reinhold, Luigi Frunzio, S.~M. Girvin, Liang Jiang, M.~H. Devoret, and R.~J. Schoelkopf.
\newblock {On-demand quantum state transfer and entanglement between remote microwave cavity memories}.
\newblock {\em Nat. Phys.}, 14:705--710, July 2018.

\bibitem{azuma2015all}
Koji Azuma, Kiyoshi Tamaki, and Hoi-Kwong Lo.
\newblock All-photonic quantum repeaters.
\newblock {\em Nature communications}, 6(1):6787, 2015.

\bibitem{blinov2021comparison}
Sergey Blinov, B~Wu, and C~Monroe.
\newblock Comparison of cloud-based ion trap and superconducting quantum computer architectures.
\newblock {\em AVS Quantum Science}, 3(3), 2021.

\bibitem{Bravyi2023Aug}
Sergey Bravyi, Andrew~W. Cross, Jay~M. Gambetta, Dmitri Maslov, Patrick Rall, and Theodore~J. Yoder.
\newblock {High-threshold and low-overhead fault-tolerant quantum memory}.
\newblock {\em arXiv}, August 2023.

\bibitem{bravyi2023high}
Sergey Bravyi, Andrew~W Cross, Jay~M Gambetta, Dmitri Maslov, Patrick Rall, and Theodore~J Yoder.
\newblock High-threshold and low-overhead fault-tolerant quantum memory.
\newblock {\em arXiv preprint arXiv:2308.07915}, 2023.

\bibitem{bravyi2016trading}
Sergey Bravyi, Graeme Smith, and John~A Smolin.
\newblock Trading classical and quantum computational resources.
\newblock {\em Physical Review X}, 6(2):021043, 2016.

\bibitem{Burkhart2021Aug}
Luke~D. Burkhart, James~D. Teoh, Yaxing Zhang, Christopher~J. Axline, Luigi Frunzio, M.~H. Devoret, Liang Jiang, S.~M. Girvin, and R.~J. Schoelkopf.
\newblock {Error-Detected State Transfer and Entanglement in a Superconducting Quantum Network}.
\newblock {\em PRX Quantum}, 2(3):030321, August 2021.

\bibitem{Campagne-Ibarcq2018May}
P.~Campagne-Ibarcq, E.~Zalys-Geller, A.~Narla, S.~Shankar, P.~Reinhold, L.~Burkhart, C.~Axline, W.~Pfaff, L.~Frunzio, R.~J. Schoelkopf, and M.~H. Devoret.
\newblock {Deterministic Remote Entanglement of Superconducting Circuits through Microwave Two-Photon Transitions}.
\newblock {\em Phys. Rev. Lett.}, 120(20):200501, May 2018.

\bibitem{coopmans2021}
Tim Coopmans, Robert Knegjens, Axel Dahlberg, David Maier, Loek Nijsten, Julio {de Oliveira Filho}, Martijn Papendrecht, Julian Rabbie, Filip Rozp{\k{e}}dek, Matthew Skrzypczyk, Leon Wubben, Walter {de Jong}, Damian Podareanu, Ariana {Torres-Knoop}, David Elkouss, and Stephanie Wehner.
\newblock {{NetSquid}}, a {{NETwork Simulator}} for {{QUantum Information}} using {{Discrete}} events.
\newblock {\em Communications Physics}, 4(1):1--15, July 2021.

\bibitem{dai2021}
Wenhan Dai, Anthony Rinaldi, and Don Towsley.
\newblock Entanglement {{Swapping}} in {{Quantum Switches}}: {{Protocol Design}} and {{Stability Analysis}}.
\newblock {\em arXiv:2110.04116 [quant-ph]}, October 2021.

\bibitem{eisert2000}
J.~Eisert, K.~Jacobs, P.~Papadopoulos, and M.~B. Plenio.
\newblock Optimal local implementation of nonlocal quantum gates.
\newblock {\em Physical Review A}, 62(5):052317, October 2000.

\bibitem{Fowler2012Oct}
Austin~G. Fowler.
\newblock {Time-optimal quantum computation}.
\newblock {\em arXiv}, October 2012.

\bibitem{Fowler2018Aug}
Austin~G. Fowler and Craig Gidney.
\newblock {Low overhead quantum computation using lattice surgery}.
\newblock {\em arXiv}, August 2018.

\bibitem{Fowler2012Sep}
Austin~G. Fowler, Matteo Mariantoni, John~M. Martinis, and Andrew~N. Cleland.
\newblock {Surface codes: Towards practical large-scale quantum computation}.
\newblock {\em Phys. Rev. A}, 86(3):032324, September 2012.

\bibitem{Fowler2010May}
Austin~G. Fowler, David~S. Wang, Charles~D. Hill, Thaddeus~D. Ladd, Rodney Van~Meter, and Lloyd C.~L. Hollenberg.
\newblock {Surface Code Quantum Communication}.
\newblock {\em Phys. Rev. Lett.}, 104(18):180503, May 2010.

\bibitem{Gao2021Nov}
Yvonne~Y. Gao, M.~Adriaan Rol, Steven Touzard, and Chen Wang.
\newblock {Practical Guide for Building Superconducting Quantum Devices}.
\newblock {\em PRX Quantum}, 2(4):040202, November 2021.

\bibitem{Gidney2021Jul}
Craig Gidney.
\newblock {Stim: a fast stabilizer circuit simulator}.
\newblock {\em Quantum}, 5:497, July 2021.

\bibitem{Grigoras2022Sep}
K.~Grigoras, N.~Yurttag{\ifmmode\ddot{u}\else\"{u}\fi}l, J.-P. Kaikkonen, E.~T. Mannila, P.~Eskelinen, D.~P. Lozano, H.-X. Li, M.~Rommel, D.~Shiri, N.~Tiencken, S.~Simbierowicz, A.~Ronzani, J.~H{\ifmmode\ddot{a}\else\"{a}\fi}tinen, D.~Datta, V.~Vesterinen, L.~Gr{\ifmmode\ddot{o}\else\"{o}\fi}nberg, J.~Bizn{\ifmmode\acute{a}\else\'{a}\fi}rov{\ifmmode\acute{a}\else\'{a}\fi}, A.~Fadavi Roudsari, S.~Kosen, A.~Osman, M.~Prunnila, J.~Hassel, J.~Bylander, and J.~Govenius.
\newblock {Qubit-Compatible Substrates With Superconducting Through-Silicon Vias}.
\newblock {\em IEEE Transactions on Quantum Engineering}, 3:ArticleSequenceNumber:5100310, September 2022.

\bibitem{hahn2018}
F.~Hahn, A.~Pappa, and J.~Eisert.
\newblock Quantum network routing and local complementation.
\newblock {\em npj Quantum Information}, 5(1):1--7, September 2019.

\bibitem{hein2006}
M.~Hein, W.~D{\"u}r, J.~Eisert, R.~Raussendorf, M.~{Van den Nest}, and H.-J. Briegel.
\newblock Entanglement in graph states and its applications.
\newblock {\em Quantum Computers, Algorithms and Chaos}, pages 115--218, 2006.

\bibitem{Heinsoo2018Sep}
Johannes Heinsoo, Christian~Kraglund Andersen, Ants Remm, Sebastian Krinner, Theodore Walter, Yves Salath{\ifmmode\acute{e}\else\'{e}\fi}, Simone Gasparinetti, Jean-Claude Besse, Anton Poto{\ifmmode\check{c}\else\v{c}\fi}nik, Andreas Wallraff, and Christopher Eichler.
\newblock {Rapid High-fidelity Multiplexed Readout of Superconducting Qubits}.
\newblock {\em Phys. Rev. Appl.}, 10(3):034040, September 2018.

\bibitem{Herr2017Jan}
Daniel Herr, Franco Nori, and Simon~J. Devitt.
\newblock {Lattice surgery translation for quantum computation}.
\newblock {\em New J. Phys.}, 19(1):013034, January 2017.

\bibitem{Heyfron2018Sep}
Luke~E. Heyfron and Earl~T. Campbell.
\newblock {An efficient quantum compiler that reduces T count}.
\newblock {\em Quantum Sci. Technol.}, 4(1):015004, September 2018.

\bibitem{Higgott2022Jul}
Oscar Higgott.
\newblock {PyMatching: A Python Package for Decoding Quantum Codes with Minimum-Weight Perfect Matching}.
\newblock {\em ACM Transactions on Quantum Computing}, 3(3):1--16, July 2022.

\bibitem{Holmes2019Jun}
Adam Holmes, Yongshan Ding, Ali Javadi-Abhari, Diana Franklin, Margaret Martonosi, and Frederic~T. Chong.
\newblock {Resource optimized quantum architectures for surface code implementations of magic-state distillation}.
\newblock {\em Microprocess. Microsyst.}, 67:56--70, June 2019.

\bibitem{Horsman2012Dec}
Dominic Horsman, Austin~G. Fowler, Simon Devitt, and Rodney Van~Meter.
\newblock {Surface code quantum computing by lattice surgery}.
\newblock {\em New J. Phys.}, 14(12):123011, December 2012.

\bibitem{Huang2021Apr}
Sihao Huang, Benjamin Lienhard, Greg Calusine, Antti Veps{\ifmmode\ddot{a}\else\"{a}\fi}l{\ifmmode\ddot{a}\else\"{a}\fi}inen, Jochen Braum{\ifmmode\ddot{u}\else\"{u}\fi}ller, David~K. Kim, Alexander~J. Melville, Bethany~M. Niedzielski, Jonilyn~L. Yoder, Bharath Kannan, Terry~P. Orlando, Simon Gustavsson, and William~D. Oliver.
\newblock {Microwave Package Design for Superconducting Quantum Processors}.
\newblock {\em PRX Quantum}, 2(2):020306, April 2021.

\bibitem{IBMRoadmap2023Nov}
{IBM Quantum Computing {$\vert$} Roadmap}, November 2023.
\newblock [Online; accessed 15. Nov. 2023].

\bibitem{jung2020ruche}
Dai~Cheol Jung, Scott Davidson, Chun Zhao, Dustin Richmond, and Michael~Bedford Taylor.
\newblock Ruche networks: Wire-maximal, no-fuss nocs: Special session paper.
\newblock In {\em 2020 14th IEEE/ACM International Symposium on Networks-on-Chip (NOCS)}, pages 1--8. IEEE, 2020.

\bibitem{Kitaev2003Jan}
A.~{\relax Yu}. Kitaev.
\newblock {Fault-tolerant quantum computation by anyons}.
\newblock {\em Ann. Phys.}, 303(1):2--30, January 2003.

\bibitem{Krinner2019Dec}
S.~Krinner, S.~Storz, P.~Kurpiers, P.~Magnard, J.~Heinsoo, R.~Keller, J.~L{\ifmmode\ddot{u}\else\"{u}\fi}tolf, C.~Eichler, and A.~Wallraff.
\newblock {Engineering cryogenic setups for 100-qubit scale superconducting circuit systems}.
\newblock {\em EPJ Quantum Technol.}, 6(1):1--29, December 2019.

\bibitem{Kurpiers2018Jun}
P.~Kurpiers, P.~Magnard, T.~Walter, B.~Royer, M.~Pechal, J.~Heinsoo, Y.~Salath{\ifmmode\acute{e}\else\'{e}\fi}, A.~Akin, S.~Storz, J.-C. Besse, S.~Gasparinetti, A.~Blais, and A.~Wallraff.
\newblock {Deterministic quantum state transfer and remote entanglement using microwave photons}.
\newblock {\em Nature}, 558:264--267, June 2018.

\bibitem{Lao2018Sep}
L.~Lao, B.~van Wee, I.~Ashraf, J.~van Someren, N.~Khammassi, K.~Bertels, and C.~G. Almudever.
\newblock {Mapping of lattice surgery-based quantum circuits on surface code architectures}.
\newblock {\em Quantum Sci. Technol.}, 4(1):015005, September 2018.

\bibitem{Leung2019Feb}
N.~Leung, Y.~Lu, S.~Chakram, R.~K. Naik, N.~Earnest, R.~Ma, K.~Jacobs, A.~N. Cleland, and D.~I. Schuster.
\newblock {Deterministic bidirectional communication and remote entanglement generation between superconducting qubits}.
\newblock {\em npj Quantum Inf.}, 5(18):1--5, February 2019.

\bibitem{li2023all}
Chen-Long Li, Yao Fu, Wen-Bo Liu, Yuan-Mei Xie, Bing-Hong Li, Min-Gang Zhou, Hua-Lei Yin, and Zeng-Bing Chen.
\newblock All-photonic quantum repeater for multipartite entanglement generation.
\newblock {\em Optics Letters}, 48(5):1244--1247, 2023.

\bibitem{li2019experimental}
Zheng-Da Li, Rui Zhang, Xu-Fei Yin, Li-Zheng Liu, Yi~Hu, Yu-Qiang Fang, Yue-Yang Fei, Xiao Jiang, Jun Zhang, Li~Li, et~al.
\newblock Experimental quantum repeater without quantum memory.
\newblock {\em Nature photonics}, 13(9):644--648, 2019.

\bibitem{Litinski2019Mar}
Daniel Litinski.
\newblock {A Game of Surface Codes: Large-Scale Quantum Computing with Lattice Surgery}.
\newblock {\em Quantum}, 3:128, March 2019.

\bibitem{Magnard2020Dec}
P.~Magnard, S.~Storz, P.~Kurpiers, J.~Sch{\ifmmode\ddot{a}\else\"{a}\fi}r, F.~Marxer, J.~L{\ifmmode\ddot{u}\else\"{u}\fi}tolf, T.~Walter, J.-C. Besse, M.~Gabureac, K.~Reuer, A.~Akin, B.~Royer, A.~Blais, and A.~Wallraff.
\newblock {Microwave Quantum Link between Superconducting Circuits Housed in Spatially Separated Cryogenic Systems}.
\newblock {\em Phys. Rev. Lett.}, 125(26):260502, December 2020.

\bibitem{moller2017impact}
Matthias M{\"o}ller and Cornelis Vuik.
\newblock On the impact of quantum computing technology on future developments in high-performance scientific computing.
\newblock {\em Ethics and information technology}, 19:253--269, 2017.

\bibitem{Mosca2021Oct}
Michele Mosca and Priyanka Mukhopadhyay.
\newblock {A polynomial time and space heuristic algorithm for T-count}.
\newblock {\em Quantum Sci. Technol.}, 7(1):015003, October 2021.

\bibitem{Niu2023Mar}
Jingjing Niu, Libo Zhang, Yang Liu, Jiawei Qiu, Wenhui Huang, Jiaxiang Huang, Hao Jia, Jiawei Liu, Ziyu Tao, Weiwei Wei, Yuxuan Zhou, Wanjing Zou, Yuanzhen Chen, Xiaowei Deng, Xiuhao Deng, Changkang Hu, Ling Hu, Jian Li, Dian Tan, Yuan Xu, Fei Yan, Tongxing Yan, Song Liu, Youpeng Zhong, Andrew~N. Cleland, and Dapeng Yu.
\newblock {Low-loss interconnects for modular superconducting quantum processors}.
\newblock {\em Nat. Electron.}, 6:235--241, March 2023.

\bibitem{national2019quantum}
National~Academies of~Sciences~Engineering and Medicine \and others.
\newblock Quantum computing: progress and prospects.
\newblock 2019.

\bibitem{Paler2017Apr}
Alexandru Paler, Ilia Polian, Kae Nemoto, and Simon~J. Devitt.
\newblock {Fault-tolerant, high-level quantum circuits: form, compilation and description}.
\newblock {\em Quantum Sci. Technol.}, 2(2):025003, April 2017.

\bibitem{panigrahy2022}
Nitish~K. Panigrahy, Thirupathaiah Vasantam, Don Towsley, and Leandros Tassiulas.
\newblock On the {{Capacity Region}} of a {{Quantum Switch}} with {{Entanglement Purification}}, December 2022.

\bibitem{pirandola2019}
Stefano Pirandola.
\newblock End-to-end capacities of a quantum communication network.
\newblock {\em Communications Physics}, 2(1):1--10, May 2019.

\bibitem{Place2021Mar}
Alexander P.~M. Place, Lila V.~H. Rodgers, Pranav Mundada, Basil~M. Smitham, Mattias Fitzpatrick, Zhaoqi Leng, Anjali Premkumar, Jacob Bryon, Andrei Vrajitoarea, Sara Sussman, Guangming Cheng, Trisha Madhavan, Harshvardhan~K. Babla, Xuan~Hoang Le, Youqi Gang, Berthold J{\ifmmode\ddot{a}\else\"{a}\fi}ck, Andr{\ifmmode\acute{a}\else\'{a}\fi}s Gyenis, Nan Yao, Robert~J. Cava, Nathalie~P. de~Leon, and Andrew~A. Houck.
\newblock {New material platform for superconducting transmon qubits with coherence times exceeding 0.3 milliseconds}.
\newblock {\em Nat. Commun.}, 12(1779):1--6, March 2021.

\bibitem{Preskill1998Jan}
John Preskill.
\newblock {Reliable quantum computers}.
\newblock {\em Proc. R. Soc. Lond. A.}, 454(1969):385--410, January 1998.

\bibitem{Qiskit}
{Qiskit contributors}.
\newblock Qiskit: An open-source framework for quantum computing, 2023.

\bibitem{Foxen2020Sep}
Google A.~I. Quantum, B.~Foxen, C.~Neill, A.~Dunsworth, P.~Roushan, B.~Chiaro, A.~Megrant, J.~Kelly, Zijun Chen, K.~Satzinger, R.~Barends, F.~Arute, K.~Arya, R.~Babbush, D.~Bacon, J.~C. Bardin, S.~Boixo, D.~Buell, B.~Burkett, Yu~Chen, R.~Collins, E.~Farhi, A.~Fowler, C.~Gidney, M.~Giustina, R.~Graff, M.~Harrigan, T.~Huang, S.~V. Isakov, E.~Jeffrey, Z.~Jiang, D.~Kafri, K.~Kechedzhi, P.~Klimov, A.~Korotkov, F.~Kostritsa, D.~Landhuis, E.~Lucero, J.~McClean, M.~McEwen, X.~Mi, M.~Mohseni, J.~Y. Mutus, O.~Naaman, M.~Neeley, M.~Niu, A.~Petukhov, C.~Quintana, N.~Rubin, D.~Sank, V.~Smelyanskiy, A.~Vainsencher, T.~C. White, Z.~Yao, P.~Yeh, A.~Zalcman, H.~Neven, and J.~M. Martinis.
\newblock {Demonstrating a Continuous Set of Two-Qubit Gates for Near-Term Quantum Algorithms}.
\newblock {\em Phys. Rev. Lett.}, 125(12):120504, September 2020.

\bibitem{Ramette2023Feb}
Joshua Ramette, Josiah Sinclair, Nikolas~P. Breuckmann, and Vladan Vuleti{\ifmmode\acute{c}\else\'{c}\fi}.
\newblock {Fault-Tolerant Connection of Error-Corrected Qubits with Noisy Links}.
\newblock {\em arXiv}, February 2023.

\bibitem{reiserer2015cavity}
Andreas Reiserer and Gerhard Rempe.
\newblock Cavity-based quantum networks with single atoms and optical photons.
\newblock {\em Reviews of Modern Physics}, 87(4):1379, 2015.

\bibitem{smith2022scaling}
Kaitlin~N Smith, Gokul~Subramanian Ravi, Jonathan~M Baker, and Frederic~T Chong.
\newblock Scaling superconducting quantum computers with chiplet architectures.
\newblock In {\em 2022 55th IEEE/ACM International Symposium on Microarchitecture (MICRO)}, pages 1092--1109. IEEE, 2022.

\bibitem{Stehlik2021Aug}
J.~Stehlik, D.~M. Zajac, D.~L. Underwood, T.~Phung, J.~Blair, S.~Carnevale, D.~Klaus, G.~A. Keefe, A.~Carniol, M.~Kumph, Matthias Steffen, and O.~E. Dial.
\newblock {Tunable Coupling Architecture for Fixed-Frequency Transmon Superconducting Qubits}.
\newblock {\em Phys. Rev. Lett.}, 127(8):080505, August 2021.

\bibitem{stein2023microarchitectures}
Samuel Stein, Sara Sussman, Teague Tomesh, Charles Guinn, Esin Tureci, Sophia~Fuhui Lin, Wei Tang, James Ang, Srivatsan Chakram, Ang Li, et~al.
\newblock Microarchitectures for heterogeneous superconducting quantum computers.
\newblock {\em arXiv preprint arXiv:2305.03243}, 2023.

\bibitem{vardoyan2021a}
Gayane Vardoyan, Saikat Guha, Philippe Nain, and Don Towsley.
\newblock On the {{Stochastic Analysis}} of a {{Quantum Entanglement Distribution Switch}}.
\newblock {\em IEEE Transactions on Quantum Engineering}, 2:1--16, 2021.

\bibitem{Wang2015Oct}
C.~Wang, C.~Axline, Y.~Y. Gao, T.~Brecht, Y.~Chu, L.~Frunzio, M.~H. Devoret, and R.~J. Schoelkopf.
\newblock {Surface participation and dielectric loss in superconducting qubits}.
\newblock {\em Appl. Phys. Lett.}, 107(16), October 2015.

\bibitem{Wang2022Jan}
Chenlu Wang, Xuegang Li, Huikai Xu, Zhiyuan Li, Junhua Wang, Zhen Yang, Zhenyu Mi, Xuehui Liang, Tang Su, Chuhong Yang, Guangyue Wang, Wenyan Wang, Yongchao Li, Mo~Chen, Chengyao Li, Kehuan Linghu, Jiaxiu Han, Yingshan Zhang, Yulong Feng, Yu~Song, Teng Ma, Jingning Zhang, Ruixia Wang, Peng Zhao, Weiyang Liu, Guangming Xue, Yirong Jin, and Haifeng Yu.
\newblock {Towards practical quantum computers: transmon qubit with a lifetime approaching 0.5 milliseconds}.
\newblock {\em npj Quantum Inf.}, 8(3):1--6, January 2022.

\bibitem{wu2022collcomm}
Anbang Wu, Yufei Ding, and Ang Li.
\newblock Collcomm: Enabling efficient collective quantum communication based on epr buffering.
\newblock {\em arXiv preprint arXiv:2208.06724}, 2022.

\bibitem{wu2022autocomm}
Anbang Wu, Hezi Zhang, Gushu Li, Alireza Shabani, Yuan Xie, and Yufei Ding.
\newblock Autocomm: A framework for enabling efficient communication in distributed quantum programs.
\newblock In {\em 2022 55th IEEE/ACM International Symposium on Microarchitecture (MICRO)}, pages 1027--1041. IEEE, 2022.

\bibitem{Wu2023May}
Yue Wu and Lin Zhong.
\newblock {Fusion Blossom: Fast MWPM Decoders for QEC}.
\newblock {\em arXiv}, May 2023.

\bibitem{wurtz2022industry}
Jonathan Wurtz, Pedro~LS Lopes, Nathan Gemelke, Alexander Keesling, and Shengtao Wang.
\newblock Industry applications of neutral-atom quantum computing solving independent set problems.
\newblock {\em arXiv preprint arXiv:2205.08500}, 2022.

\bibitem{Zhang2019Mar}
Fang Zhang and Jianxin Chen.
\newblock {Optimizing T gates in Clifford+T circuit as $\pi/4$ rotations around Paulis}.
\newblock {\em arXiv}, March 2019.

\bibitem{Zhong2021Feb}
Youpeng Zhong, Hung-Shen Chang, Audrey Bienfait, {\ifmmode\acute{E}\else\'{E}\fi}tienne Dumur, Ming-Han Chou, Christopher~R. Conner, Joel Grebel, Rhys~G. Povey, Haoxiong Yan, David~I. Schuster, and Andrew~N. Cleland.
\newblock {Deterministic multi-qubit entanglement in a quantum network}.
\newblock {\em Nature}, 590:571--575, February 2021.

\end{thebibliography}

\end{document}